# Quality-Net: An End-to-End Non-intrusive Speech Quality Assessment Model based on BLSTM

*Szu-Wei Fu* [12], *Yu Tsao* [1], *Hsin-Te Hwang* [3], *Hsin-Min Wang* [3]

[1] Research Center for Information Technology Innovation, Academia Sinica, Taipei, Taiwan
[2] Department of Computer Science and Information Engineering, National Taiwan University, Taipei, Taiwan
[3] Institute of Information Science, Academia Sinica, Taipei, Taiwan

{jasonfu,yu.tsao}@citi.sinica.edu.tw; {hwanght,whm}@iis.sinica.edu.tw

## Abstract

Nowadays, most of the objective speech quality assessment tools (e.g., perceptual evaluation of speech quality (PESQ)) are based on the comparison of the degraded/processed speech with its clean counterpart. The need of a "golden" reference considerably restricts the practicality of such assessment tools in real-world scenarios since the clean reference usually cannot be accessed. On the other hand, human beings can readily evaluate the speech quality without any reference (e.g., mean opinion score (MOS) tests), implying the existence of an objective and non-intrusive (no clean reference needed) quality assessment mechanism. In this study, we propose a novel end-to-end, non-intrusive speech quality evaluation model, termed Quality-Net, based on bidirectional long short-term memory. The evaluation of utterance-level quality in Quality-Net is based on the frame-level assessment. Frame constraints and sensible initializations of forget gate biases are applied to learn meaningful frame-level quality assessment from the utterance-level quality label. Experimental results show that Quality-Net can yield high correlation to PESQ (0.9 for the noisy speech and 0.84 for the speech processed by speech enhancement). We believe that Quality-Net has potential to be used in a wide variety of applications of speech signal processing.

## 1. Introduction

Speech quality is a subjective opinion, based on a listener's feeling to the heard speech. Therefore, objective assessment of speech quality is challenging especially when a clean reference does not exist (also called non-intrusive or single-ended speech quality assessment). Although the perceptual evaluation of speech quality (PESQ) [1] is widely used to evaluate the speech quality in industrial applications [2-4], the need for a "golden" reference considerably restricts the applicability of such assessment tools in real-world scenarios. For example, it is difficult to rely on PESQ to judge the noise level for automatically turning on the speech enhancement function in mobile communications or automatic speech recognition (ASR). However, human beings can readily evaluate the speech quality without any reference. In other words, the human listening perception can be treated as a mapping function to map any speech utterance to a corresponding quality score.

Although the ITU-T released recommendation P.563 as its standard algorithm for non-intrusive objective speech quality assessment, it is designed for 3.1-kHz (narrow band) telephony applications [5]. Several non-intrusive speech quality assessment models have also been proposed [6-22]. Sharma *et al.*[8] employed classification and regression trees (CART) to predict the quality score based on many handcrafted features. Soni *et al.*[22] applied a subband autoencoder to first extract features to be used by the following neural-network-based prediction model. Although these methods have already achieved good prediction results, the features (most are complex handcrafted features) used for prediction are not jointly optimized with the back-end assessment model (not end-to-end). In addition, these models are simply treated as a black box, which also restricts the possible further applications.

It has been shown that a quality assessment model can also guide another model to learn human perception [23, 24]. Specifically, Talebi *et al.*[24] applied an image assessment model as a perceptual loss in training an image enhancement model. By simultaneously maximizing the assessment score and minimizing the reconstruction loss (e.g., $L_2$ loss), the trained enhancement model can generate more appealing images. The key in successfully combining these two models is that the assessment model is also an end-to-end model and no handcrafted features were involved; therefore, the gradients can be back propagated from the perceptual loss.

Recently, deep learning has shown its strong capacity to learn a mapping function in many different applications. Herein, we propose a novel, end-to-end, and non-intrusive speech quality evaluation model, termed Quality-Net, based on bidirectional long short-term memory (BLSTM). In addition, to prevent Quality-Net from becoming an incomprehensible black box, its structure is designed to automatically learn (infer) a reasonable frame-level quality. This gives Quality-Net the ability to locate the degraded regions in an utterance. Although our ultimate goal is to learn the mapping function of the human listening perception, an off-the-shelf data set with labels that meets our requirements does not exist (here, we focus on predicting the quality of noisy speech and enhanced speech given by a deep-learning-based speech enhancement model). Therefore, we apply Quality-Net to predict the PESQ scores. The experimental results also serve as guidelines (e.g., number of data needed) for future construction of the required data set.

In our previous work [25], we have successfully optimized the short-time objective intelligibility (STOI) [26] score in training a speech enhancement model. Although the model can be readily optimized by any differentiable metrics, some functions in PESQ computation are non-continuous; therefore, the gradient-descent-based optimization cannot be directly used. By contrast, since Quality-Net is an end-to-end assessment model, it can be combined with a speech enhancement model to boost the PESQ score of enhanced speech. This will be an important future work of this study. To our best knowledge, Quality-Net is the first end-to-end, and non-intrusive quality assessment model to yield frame-level quality.

## 2. Quality-Net

The goal of this paper is to propose a non-intrusive speech quality assessment model. As speech utterances have different lengths, this model has to map $u \in R^{T(u)}$ to $Q \in R^1$, where $u$ is the input speech, $Q$ is the estimated quality score, and $T(u)$ is the length of the input speech $u$; ($T(u)$ can be the number of sample points or number of frames in $u$ for the time-domain-waveform- or frequency-domain-spectrogram-based estimations, respectively). To overcome this mapping restriction (variable-length input, fixed-length output), BLSTM is employed to mimic a human listening perception system for quality estimation through pairs of (speech utterance, quality score) training data. This model is called Quality-Net herein, and we adopt the magnitude spectrogram as the input feature. Therefore, after reading the whole spectrogram, Quality-Net can predict a score for speech quality evaluation. In addition, to prevent Quality-Net from becoming an incomprehensible black box, its structure is designed to render the intermediate evaluation process more meaningful. Specifically, we designed Quality-Net to automatically learn (infer) a reasonable frame-level quality even though the quality label in the training data is utterance wise. Here, reasonable frame-level quality estimation means that if noise or speech distortion occurs in one frame, then its quality score should be decreased accordingly. Subsequently, the final estimated utterance-level quality score $Q$ is obtained by combining the frame-wise scores $q_t$ through a global average, as shown in Fig. 1. Note that the global average function not only solves the mapping restriction of the fixed-length output, but also provides $q_t$ the physical meaning of frame-level quality. This particular structure of Quality-Net is designed to mimic the process that people evaluate the speech quality. In the following, we introduce the detailed settings of Quality-Net to infer the reasonable frame-level quality from the utterance-level quality label.

### 2.1. Conditional constraint on frame quality assessment

As stated in the previous section, we only have the utterance-level speech quality labels. When the noise is not stationary (i.e., the frame-wise SNR is varying), or the degree of speech distortion is not the same across frames, it is not suitable to directly assign the utterance-level quality label to every individual frame within the input utterance. However, the inconsistency of utterance-level and frame-level scores becomes insignificant when the quality of the input utterance is high (e.g., if the speech quality is evaluated as perfect by human, then there should be no degradation anywhere, and each frame can be assigned with a perfect score). Based on this concept, we incorporate a conditional frame-wise constraint in the objective function of Quality-Net: when the speech quality of the input utterance is lower/higher, the frame-wise constraints will be given lower/higher weights. Accordingly, we derive the objective function for Quality-Net as:

$$O = \frac{1}{S}\sum_{s=1}^{S}[(\hat{Q}_s - Q_s)^2 + \alpha(\hat{Q}_s)\sum_{t=1}^{T(u_s)}(\hat{Q}_s - q_{s,t})^2] \quad (1)$$

where $\alpha(\hat{Q}_s)$ is the weighting factor, which is a function of the true utterance-level quality based on:

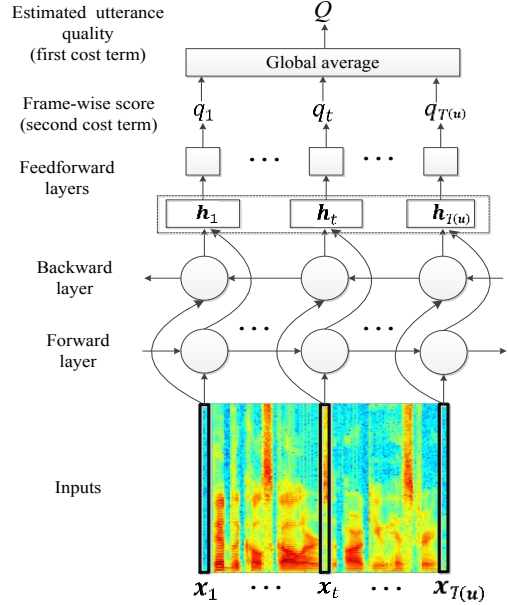

Figure 1: *Proposed Quality-Net for end-to-end, non-intrusive speech quality assessment.*

$$\alpha(\hat{Q}_s) = 10^{(\hat{Q}_s - \hat{Q}_{MAX})} \quad (2)$$

where $S$ is the total number of training utterances; $\hat{Q}_s$ and $Q_s$ are the true and estimated quality scores of the $s$-th utterance, respectively. $q_{s,t}$ is the estimated frame quality of the $t$-th frame of utterance $s$, and $\hat{Q}_{MAX}$ is the maximum quality score in the metric (e.g., $\hat{Q}_{MAX} = 5$ in MOS, and $\hat{Q}_{MAX} = 4.5$ in PESQ). Note that the first term in (1) only focuses on the accuracy of utterance-level quality and does not concern the distribution of frame-level quality. Nevertheless, the second term in (1) forces the frame-level quality to follow a uniform distribution. This constraint is more significant for speeches of higher quality as its influence exponentially decreases according to (2). In summary, for a high-quality speech, the estimated utterance-level quality consists of uniformly distributed frame-level quality with scores equal to the utterance-level quality score. This constraint also explicitly guides Quality-Net to differentiate clean frames from degraded frames.

### 2.2. Limited context influence

One of the major advantages of BLSTM is that the current estimation considers the information from the past and future contexts even though they may be several time steps away. However, the inclusion of the context information causes the frame-level quality assessment in Quality-Net to not completely focus on the condition of the current frame. Note that if only some regions of a speech utterance are contaminated by noise, then the frame-level quality scores in the clean region will also be decreased. Therefore, the comparison of frame-level quality can only hold inside an utterance, and may not be compared between different utterances. For example, given a noisy utterance $N$, if a frame, $x_{N,c}$, and another frame, $x_{N,n}$, are present in the clean and noisy regions, respectively, then $q_{N,c} > q_{N,n}$ should hold properly for Quality-Net. However, given another clean utterance $C$, if there is a clean frame, $x_{C,c}$, that has the same feature values as $x_{N,c}$ (i.e., $x_{C,c} = x_{N,c}$), owing to the context dependency of the BLSTM model, we might observe $q_{C,c} \neq q_{N,c}$ (more specifically, $q_{N,c} < q_{C,c}$ if Quality-

Net is properly trained). It is noteworthy that the estimated frame-wise quality is not actually "frame-wise," where the distant context information always influences the score. The most straightforward method to directly control the flow of context information in BLSTM is through the forget gate and recurrent weight matrix. As recommended by Jozefowicz et al. [27], to learn more long-range dependencies, the forget gate bias (Fgb) should be initialized to a larger value. Since we intend to limit the degree of context information to be used in our application, the Fgb is initialized to a smaller value, making Quality-Net prone to forget and focus more on the current frame.

## 3. Experiments

### 3.1. Experimental setup

In our experiments, the TIMIT corpus [28] was used to prepare the training and test sets. All 4620 utterances from the training set of the TIMIT database were used for training Quality-Net. These utterances were further divided into three subsets, namely clean, noisy, and enhanced sets to learn the assessment mapping function of different speech conditions. For the clean set, 250 utterances were randomly selected to keep their original clean condition. The remaining utterances were further divided into two parts to form noisy and enhanced sets. The speech utterances in the noisy set were obtained by corrupting the original speech utterances with 90 noise types, at eight SNR levels (from -10 dB to 25 dB with steps of 5 dB). For more challenging experimental conditions, each utterance was only corrupted with one noise type at one SNR level such that all the training data were unparalleled. The speech utterances in the enhanced set were corrupted with the same 90 noise types as those used in the noisy speech set, while a BLSTM-based speech enhancement model [29, 30] was applied on these utterances. The speech enhancement model was trained using 200 utterances (randomly selected from the test set of the TIMIT database) corrupted with 10 noise types, at four SNR levels (-8 dB, -4 dB, 4 dB, and 8 dB). Thus, a total of 8000 utterances were used to train the speech enhancement model. Note that the noise types used for training Quality-Net and the speech enhancement model were not overlapped, despite them coming from [31].

Another 100 utterances were randomly selected from the test set of the TIMIT database for evaluating the performance of Quality-Net on clean, noisy, and enhanced speech. For the clean test set, we used the original clean speech utterances; for the noisy test set, the 100 utterances were corrupted with four unseen noise types (engine, white, street, and baby cry), at six SNR levels (-6 dB, 0 dB, 6 dB, 12 dB, 18 dB, and 24 dB); for the enhanced set, the utterances in the noisy set were enhanced by the enhancement model above. In summary, there are total of 4900 utterances in the test set.

Quality-Net has one bidirectional LSTM layer with 100 nodes, followed by two fully connected layers, each with 50 exponential linear unit (ELU) [32] nodes and one linear node (for frame-level quality assessment). The last layer is the global average layer for obtaining the utterance-level quality score $Q$. The parameters are trained with RMSprop [33], which is a suitable optimizer for RNNs.

The quality label in the experiments was based on PESQ, and the findings can be treated as a pilot study for the future work on the assessment of MOS. Note that since Quality-Net is not based on the formulation of PESQ, its structure can also be applied to predict MOS.

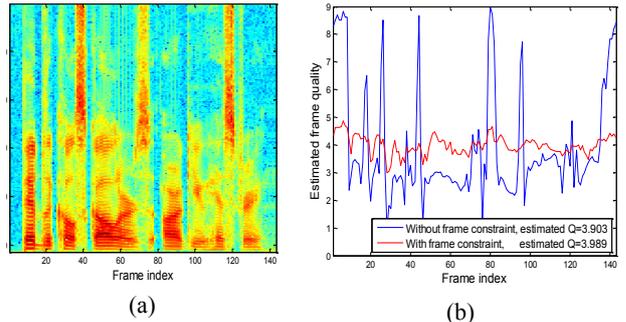

Figure 2: *Example of (a) clean speech, and (b) its corresponding frame-level quality assessment by Quality-Net, with and without frame constraint during training*.

Table 1: *Effects of frame-level quality constraint*.

|  | MSE | LCC | SRCC | Variance of frame quality in clean speech |
|---|---|---|---|---|
| without constraint | 0.1441 | 0.8559 | 0.8607 | 4.1128 |
| with constraint | **0.1266** | **0.8749** | **0.8807** | **0.2468** |

To evaluate the performance of Quality-Net, the mean square error (MSE), linear correlation coefficient (LCC), and Spearman's rank correlation coefficient (SRCC) were computed between the predicted and true PESQ scores. In the following, we first show the effects of the conditional constraint and limited context influence introduced in section 2.

### 3.2. Effects of frame-level quality constraint

We first show the effects of applying the frame quality constraint in Quality-Net training. The results of "without constraint" and "with constraint" are shown in Table 1. Note that for a model without frame constraint, it does not consider the inferred results of frame quality, and its objective function is (1) with $\alpha(\hat{Q}_s) = 0$. The results in Table 1 show that the frame constraint can effectively reduce the variance of frame quality assessment in clean speech (average over the clean utterances in the test set). In addition, since this constraint can explicitly guide Quality-Net to differentiate clean frames from degraded frames, the overall performance is also significantly improved. Fig. 2 shows an example of a clean speech utterance and its corresponding frame-level quality assessment by Quality-Net. From the figure, it is clear that application the frame-level quality constraint results in a smaller variance in the predicted frame scores.

### 3.3. Effects of initialization on forget gate bias

Next, we investigate the effects of initializing the forget gate bias in Quality-Net. The results of using different Fgb values are listed in Table 2. As shown, we can first observe that the initialization of Fgb does not significantly impact the utterance-level assessment. For further investigation on the frame-level assessment, Fig. 3 presents an example of a clean speech, a partial noisy (the 40th–100th frames, marked in black-dashed rectangle in Fig. 3(b)) utterance, and the corresponding assessment results by Quality-Net with different initial values of Fgb (Fgb= 1 and Fgb= -3). As shown in Fig. 3 (c) and (d), although both models can successfully detect the noisy region and yield low quality scores in the 40th–100th frames, the scores in the clean region (the 1st–39th and 101th–140th frames) are also decreased in Fig. 3 (c). This is because Fgb is initialized to a large value such that Quality-Net fails to forget the distant context information. Consequently, the noise in the noisy region affects the assessment of distant clean frames. This somewhat violates the desired property of frame assessment. Based on the findings in Table 2, we use the optimal setup (Fgb is initialized as -3) in the following experiments.

Table 2: *Effects of initialization on forget gate bias.*

| Initialization of Fgb | MSE | LCC | SRCC |
|---|---|---|---|
| Fgb= 1 [27] | 0.1273 | 0.8737 | 0.8721 |
| Fgb= -1 | 0.1366 | 0.8667 | 0.8731 |
| **Fgb= -3** | **0.1266** | **0.8749** | **0.8807** |
| Fgb= -5 | 0.1328 | 0.8637 | 0.8668 |

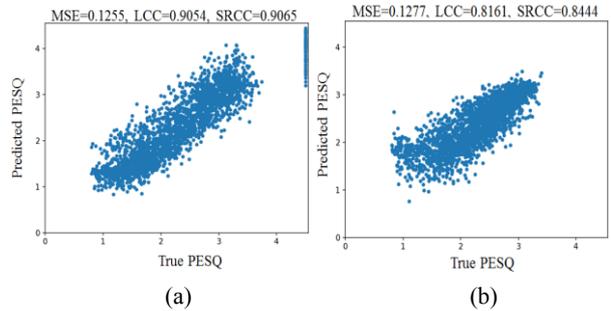

(a)          (b)

Figure 4: *Scatter plots for speech quality assessment by Quality-Net. (a) noisy and clean speech; (b) enhanced speech.*

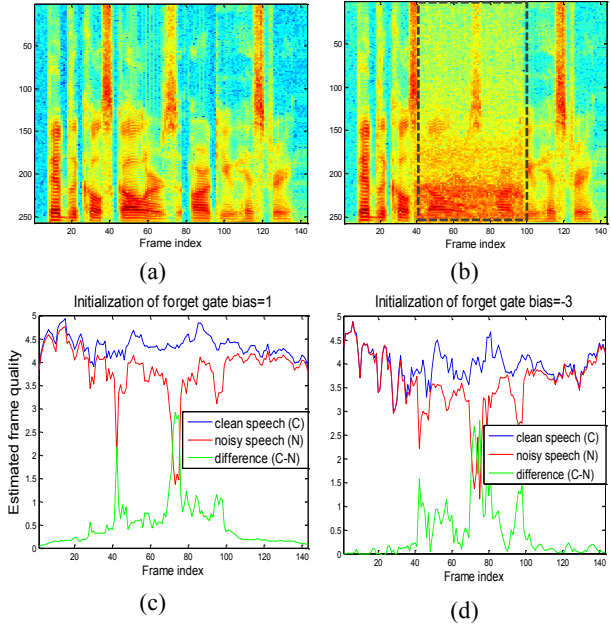

Figure 3: *Example of (a) clean speech, (b) partial noisy (street noise, the 40th–100th frames) speech, (c) assessment results by Quality-Net with the forget gate bias initialized as 1, and (d) assessment results by Quality-Net with the forget gate bias initialized as -3.*

### 3.4. Detailed assessment results

In this section, we show the detailed assessment results of Quality-Net for noisy, clean, and enhanced speech in the test set. Fig. 4(a) presents the scatter plots with the corresponding metrics for clean (the points in the upper right part) and noisy speech utterances. The assessment results for enhanced speech utterances are shown in Fig. 4(b). From these two figures, we observed that the enhanced utterances are more difficult to evaluate than the noisy and clean ones, especially for the low PESQ cases (the correlation coefficient of Fig. 4(b) is lower than that of Fig. 4(a)). Next, we compare Quality-Net with an existing two-stage model, which uses an autoencoder for feature extraction and neural network for assessment [22]. The results of the two-stage approach and Quality-Net are listed in Table 3. As shown, Quality-Net outperforms the two-stage model, possibly because Quality-Net jointly optimizes the feature extraction and assessment model.

Table 3: *Results of Quality-Net and the two-stage model.*

|  | MSE | LCC | SRCC |
|---|---|---|---|
| Autoencoder +NN [22] | 0.1529 | 0.8434 | 0.8675 |
| Quality-Net | **0.1266** | **0.8749** | **0.8807** |

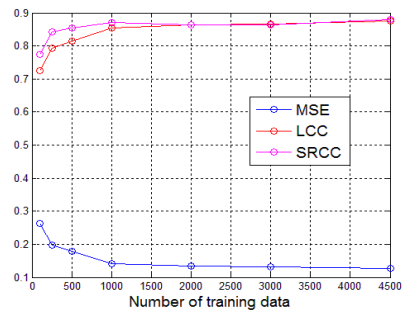

Figure 5: *Relation between the number of training utterances and the assessment results.*

### 3.5. Relation between number of training utterances and assessment performance

Although this paper focuses on predicting the PESQ scores of speech utterances, it also provides guidelines for the corpus collection, which is one of our future works. Collection of a large number of speech utterances and the corresponding MOS labels is laborious and time consuming. Therefore, we also investigated the relation between the number of training utterances and the assessment results. From Fig. 5, we observed that 1000 utterances are sufficient (the performance starts to saturate) for training Quality-Net to achieve an accurate quality prediction. Unexpectedly, the correlation coefficient of 0.7 is achieved by only 100 training utterances.

## 4. Conclusions

This paper proposed a novel, non-intrusive, and end-to-end speech quality evaluation model: Quality-Net. Our experimental results show that Quality-Net can yield a high correlation to PESQ. The non-intrusive property and frame-level quality assessment of Quality-Net considerably increases its practicality in different applications. The end-to-end framework further allows Quality-Net to be directly combined with a speech enhancement model (e.g., as a perceptual loss). Our future work includes applying Quality-Net for the assessment of MOS and employing it for speech enhancement. Through Quality-Net, we anticipate that the mismatch between the objective used in training a speech enhancement model and the human perception can be effectively reduced.